\documentclass[namedreferences]{spackap}
\usepackage{times}
\usepackage{epsfig}

\def\etal{{\it et~al.\ }}
\def\aa#1#2{{\it Astr.\ Astrophys.\ }{\bf #1}, #2}

\def\apj#1#2{{\it Astrophys.\ J.\ }{\bf #1}, #2}
\def\apjl#1#2{{\it Astrophys.\ J.\ (Letters) }{\bf #1}, #2}

\def\mn#1#2{{\it Mon.\ Not.\ R.\ astr.\ Soc.\ }{\bf #1}, #2}

\def\spose#1{\hbox to 0pt{#1\hss}}
\def\lta{\mathrel{\spose{\lower 3pt\hbox{$\mathchar"218$}}
     \raise 2.0pt\hbox{$\mathchar"13C$}}}
\def\gta{\mathrel{\spose{\lower 3pt\hbox{$\mathchar"218$}}
     \raise 2.0pt\hbox{$\mathchar"13E$}}}
\def\=#1{\overline{#1}}

\def\p{\partial}

\def\phibar{\varphi_{\rm bar}}

\def\rsun{{\rm\,R_0}}

\def\deg{^\circ}             %for angular measure in degrees

\def\mum{\mu{\rm m}}         % in math mode only

\def\kms{{\rm\,km\,s^{-1}}}

\def\pc{{\rm\,pc}}
\def\kpc{{\rm\,kpc}}

\def\msun{{\rm\,M_\odot}}

\def\gyr{{\rm\,Gyr}}

\begin{document}
\begin{article}
\begin{opening}

\title{Mass distribution in our Galaxy}
\author{Ortwin \surname{Gerhard}}
\institute{Astronomisches Institut, Universit\"at Basel, Switzerland\\
email: Ortwin.Gerhard@unibas.ch}
\runningtitle{Mass distribution in our Galaxy}
\runningauthor{Ortwin E.~Gerhard} 

\begin{ao}
Ortwin Gerhard, Astronomisches Institut der Universit\"at Basel,
Venusstrasse 7, CH-4102 Binningen, Switzerland, email: Ortwin.Gerhard@unibas.ch
\end{ao}

\begin{abstract}
  This article summarizes recent work on the luminosity and mass
  distribution of the Galactic bulge and disk, and on the mass of the
  Milky Way's dark halo. A new luminosity model consistent with the
  COBE NIR data and the apparent magnitude distributions of bulge
  clump giant stars has bulge/bar length of $\simeq 3.5\kpc$, axis
  ratios of 1:(0.3-0.4):0.3, and short disk scale-length ($\simeq
  2.1\kpc$). Gas-dynamical flows in the potential of this model 
  with constant M/L fit
  the terminal velocities in $10\deg\le |l| \le 50\deg$ very well. The
  luminous mass distribution with this M/L is consistent with the
  surface density of known matter near the Sun, but still
  underpredicts the microlensing optical depth towards the bulge.
  Together, these facts argue strongly for a massive, near-maximal
  disk in our $\sim L^\ast$, Sbc spiral Galaxy.  While the outer
  rotation curve and global mass distribution are not as readily
  measured as in similar spiral galaxies, the dark halo mass estimated
  from satellite velocities is consistent with a flat rotation curve
  continuing on from the luminous mass distribution.
\end{abstract}

\end{opening}

\section{Luminosity distribution of the bulge and disk}

Because of the strong dust obscuration by the intervening disk, the
structure of the inner Galaxy is best analyzed in the NIR.  Figure 1
shows the NIR luminosity model for the Galactic bulge/bar and disk
obtained by Bissantz \& Gerhard (2002) from the COBE/DIRBE L-band
data, for bar angle $\phibar=20\deg$ wrt the Sun-Galactic Center line.
This model was found by a penalized maximum likelihood estimation
of the data, with penalty terms encouraging a four-armed logarithmic
spiral structure in the disk (Ortiz \& L\'epine 1993) and discouraging
deviations from smoothness and triaxial symmetry.  The L-band data
used had been dereddened by Spergel, Malhotra \& Blitz (1996), with a
three-dimensional model of the dust based on the COBE/DIRBE 240$\mum$
data. The L-band data rather than the K-band data were used in order
to minimize the effects of the dust.

\begin{figure}[t]
\centerline{\epsfig{file=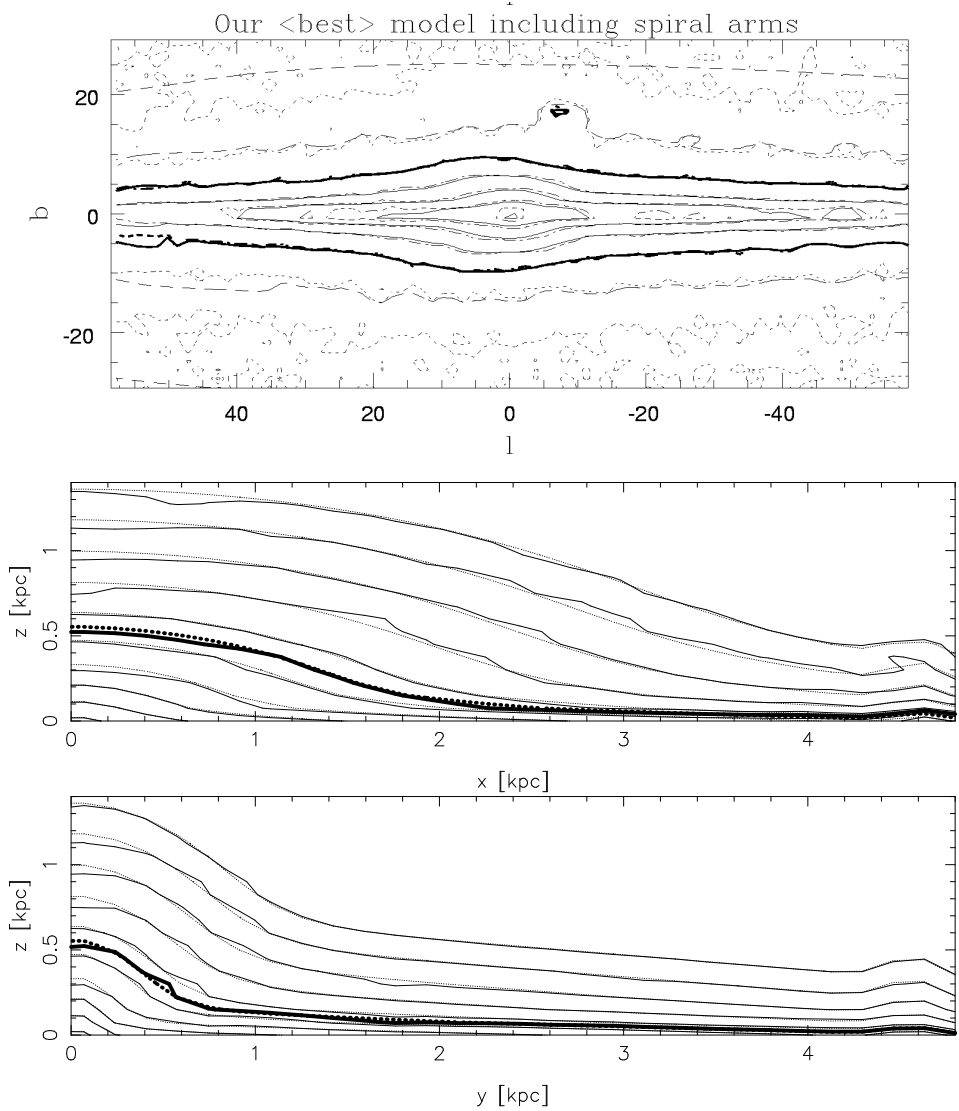,width=\hsize}}
\centerline{\epsfig{file=xz.eps,width=\hsize}}
\centerline{\epsfig{file=yz.eps,width=\hsize}}
\caption{Top: Surface brightness maps from COBE L-band data (dotted) 
  and a non-parametric luminosity model for the bulge and disk
  ($\phibar=20\deg$; dashed contours). Middle: Section through the
  model containing the bar's long and short axes.  Bottom: Same,
  containing the intermediate and short axes. Adapted from Bissantz \&
  Gerhard (2002).}
\end{figure}

\begin{figure}[t]
\centerline{\epsfig{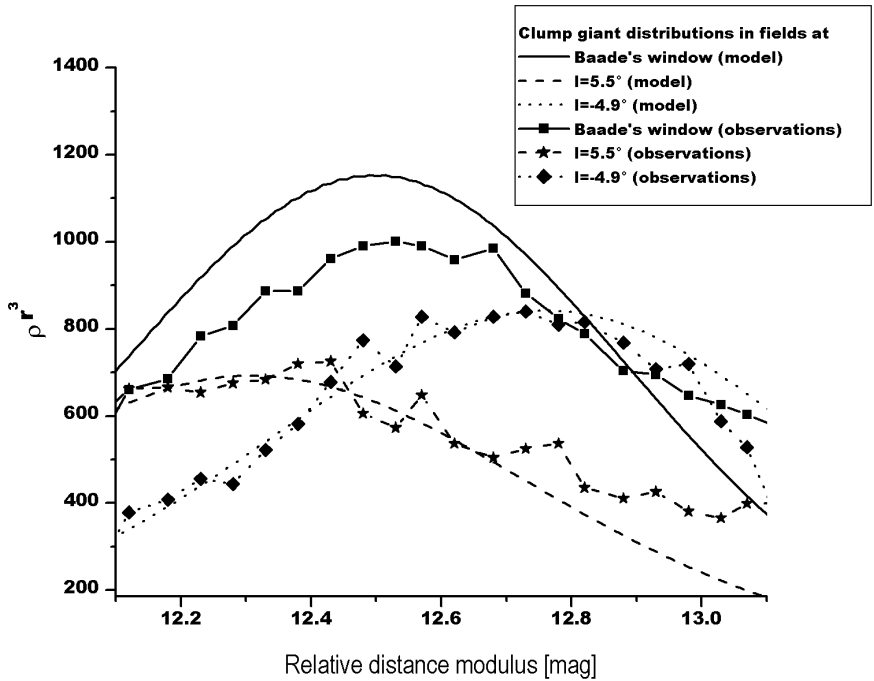}}
\caption{Apparent magnitude distributions of clump giant stars in
  three fields observed by Stanek \etal (1997). Curves show the
  predictions of the model of Fig.~1 after normalisation and
  convolution with $0.3^m$ intrinsic luminosity spread. From Bissantz
  \& Gerhard (2002).}
\end{figure}

Spiral arms make the model appear more elongated on the sky, so the
bar in Fig.~1 is more elongated than in the previous COBE model of
Binney, Gerhard \& Spergel (1997). The bar length is $\simeq 3.5\kpc$
and the axis ratios are 1:(0.3-0.4):0.3, in reasonable agreement with
starcount models (10:4:3 from clump giant stars, Stanek \etal 1997;
10:5.4:3.3 from NIR star counts in strips across the bulge,
L\'opez-Corredoira \etal 2000).  The contours in Fig.~1 show that both
the bulge and disk vertical profiles are approximately exponential.
The same is true for the radial disk profile; a fit gives a short
scale length of $\sim 2.1\kpc$.  The luminosity distribution shown in
Fig.~1 also gives a good match to the apparent magnitude distributions
of clump giants measured by Stanek \etal (1997), when an intrinisic
luminosity spread of 0.3 mag is assumed for these stars, matching well
the asymmetry of the peak positions in the fields near $l=\pm 5\deg$
wrt to Baade's window, and also the relative peak amplitudes within
$\sim10\%$ (Figure~2). This shows that the line-of-sight distribution of
luminosity in the bulge part is approximately correct. This also
constrains the bar angle; Bissantz \& Gerhard (2002) estimate
$15\deg\lta \phibar\lta 30\deg$, with the best models obtained for
$20\deg\lta \phibar\lta 25\deg$.

\section{Mass distribution of the Galactic disk}

Converting this luminosity model into a mass model requires additional
kinematic data. Bissantz, Englmaier \& Gerhard (2002) computed
hydrodynamical models in the gravitational potential of the COBE
luminosity model, assuming a constant NIR $M/L$ for the bulge and
disk, and including a dark halo component in some models. They then
fitted these models to the observed terminal velocity curve (TVC) for
HI and CO gas in the Milky Way.  Figure~3 shows the observed TVC and a
scaled model terminal curve obtained with the assumption that the
stellar component provides most of the mass in the inner Galaxy
(``maximal disk assumption'') and that $v_{\rm LSR}=220\kms$.  Due to
resolution problems in the innermost $10\deg$ the fit was restricted
to longitudes $10\deg \le |l|\le 50\deg$.  The figure shows that the
model fits  
\begin{figure}[H]
\centerline{\epsfig{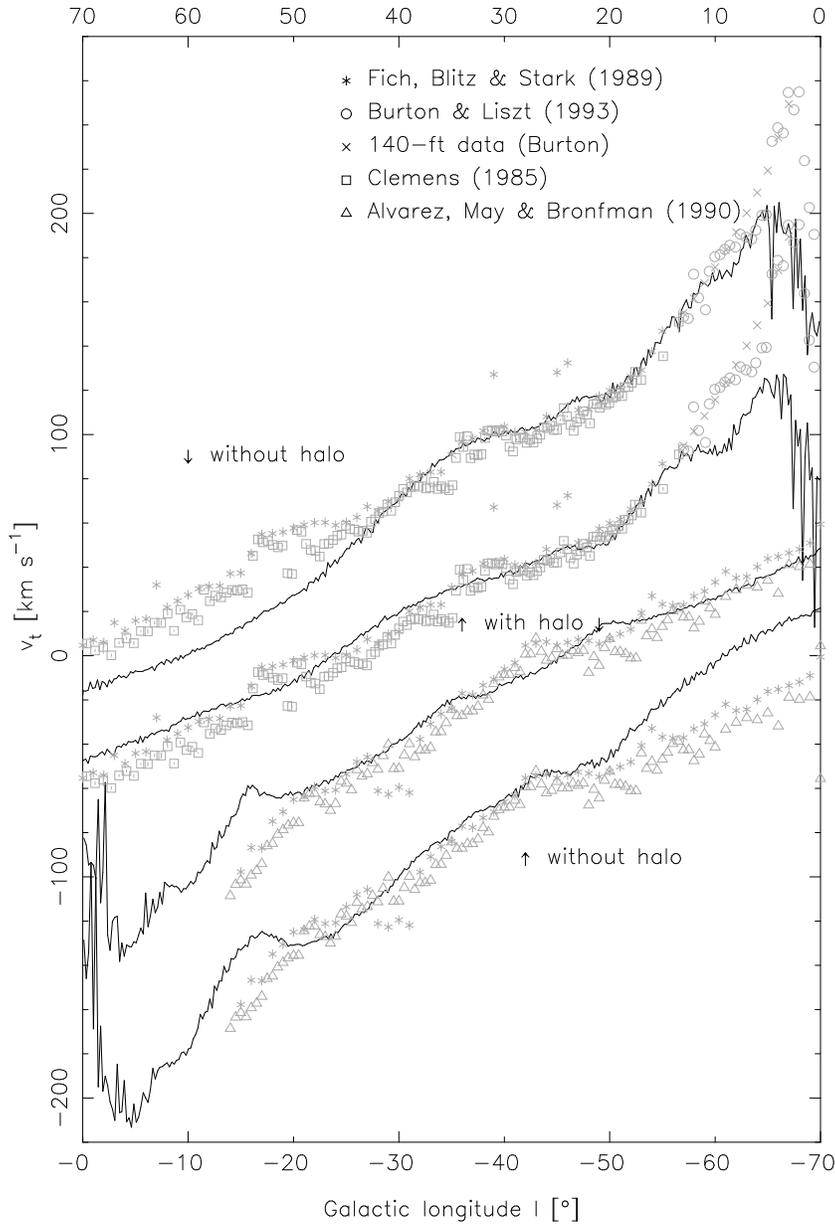}}
\caption{HI and CO terminal velocity measurements (symbols) with
  predicted curves from gas-dynamical models. The model curves are
  obtained from smooth particle hydrodynamics simulations in the
  gravitational potential of the luminosity distribution shown in
  Fig.~1, assuming constant mass-to-light ratio, with and without
  an additional quasi-isothermal dark halo component. The adopted
  LSR velocity is $220\kms$. From Bissantz \etal (2002).}
\end{figure}
\noindent
the data very well in the fitted longitude range, even including some
bumps due to the spiral arms. With this scaling, the bulge and disk
components account for just under $190\kms$ circular velocity at the
Sun. When $v_{\rm LSR}=220\kms$, an additional dark halo component is
needed to match the TVC only outside $|l|\simeq50\deg$.  Fig.~3 also
shows a model including a quasi-isothermal halo; this model extends
the good fit to the terminal velocities to where the simulation ends.
The total luminous mass in this model to $R_0=8\kpc$ is $\simeq
5.5\times 10^{10}\msun$, of which perhaps $1.3\times 10^{10}\msun$
belong to the bulge, depending on the precise decomposition.

\begin{figure}[t]
\centerline{\epsfig{file=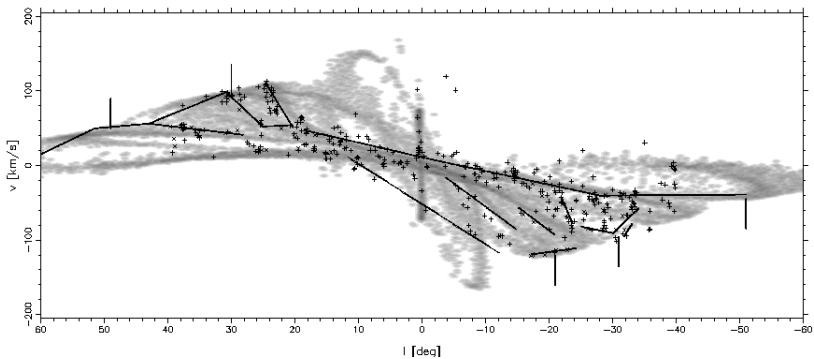,width=\hsize}}
\caption{$(l,v)$-plot for the model with halo from Fig.~3. The lines
 show spiral arm ridges in the CO data of Dame \etal (2001), while
 crosses and plus signs show molecular clouds and HII regions from
 various sources; see Bissantz \etal (2001) for references. The 
 vertical lines show the observed tangent
 point directions. From Bissantz \etal (2002).}
\end{figure}

The most important parameter in these models, the bar pattern speed,
is $\Omega_p\simeq 60\gyr^{-1}$. The spiral arms are assumed to rotate
at a second, different pattern speed, $\Omega_{sp} =20\gyr^{-1}$. An
$(l,v)$--diagram for this model with a dark halo component included is
shown in Figure 4 at the optimal phase between bar and spiral arms for
an observer at $R_0=8\kpc$ and $\phibar=20\deg$ from the bar's major
axis. The overplotted line segments represent high emission ridges
(spiral arms) in the CO data of Dame, Hartmann \& Thaddeus (2001). The
model represents the arm ridges outside corotation very well, and also
the gap in the CO $(l,v)$--diagram seen immediately outside the 3 kpc
arm.  This gap signifies that the main spiral arms remain well-defined
through the corotation region of the bar, the principal reason for
preferring a model with two pattern speeds over one with a single
corotating pattern. The model does not match well the amplitude of the
non-circular velocities of the $3\kpc$--arm. However, the transition
between the bar and spiral arms in the luminosity model is not very
well constrained by the NIR data and hence the detailed gravitational
potential in this region is likewise uncertain.  By increasing the
spiral arm amplitude in this region it is possible to obtain models
with $\simeq 50\kms$ non-circular velocity at $l=0$.  By comparing a
number of gas-dynamical models to the CO data, we have estimated the
bar pattern speed $\Omega_p=60\pm5\gyr^{-1}$, the corresponding
corotation radius $R_{\rm cr}=3.4^{+0.4}_{-0.2}\kpc$, and the
preferred bar angle $20\deg\lta \phibar\lta 25\deg$; i.e., the
Galactic bar is a fast rotator.

Matching these maximum disk models for $R_0=8\kpc$, $v_{\rm LSR}=220\kms$,
$\phibar=20-25\deg$ to the terminal velocity curve predicts
surface mass densities $\Sigma_\odot = 38$--$43 \msun/\pc^2$ at the
position of the Sun.  Azimuthally averaging over the spiral arm model
along a ring at $\rsun=8\kpc$ gives slightly higher values of
$42$--$47 \msun/\pc^2$, showing that the precise location of the Sun
between spiral arms does not have a large effect.  
For comparison, the local surface density of `identified
matter' is $48\pm 9 \msun/\pc^2$ (Kuijken \& Gilmore 1991, Flynn \&
Fuchs 1994, Holmberg \& Flynn 2000).  Of this about $23\msun/\pc^2$ is
in gas and brown and white dwarfs, which contribute most to the
uncertainty.  That the observed and predicted surface density
approximately agree lends support to the conclusion that the Galaxy
indeed has a near--maximum disk; the combined observational and model
uncertainty is about $25\%$ in mass.  By contrast, a constant-M/L NIR
Galaxy 
accounting for only $60\%$ of the rotation velocity at $2R_D$, as
advocated for spiral galaxies by Courteau \& Rix (1999), would have
$\Sigma_\odot \sim 15 \msun/\pc^2$. Turned around, if the NIR disk
and bulge are given the $\Upsilon_L$ value implied by the local
surface density measurement, then they account for the observed
terminal velocities in the inner Galaxy.  Compared to earlier
analyses, the main difference is the short disk scale--length (see
Sackett 1997; in the model above a fit in the inner Galaxy gives
$\simeq 2.1\kpc$) -- for constant $\Upsilon_L$ the Sun is well beyond
the maximum in the rotation curve from only NIR luminous matter.

The agreement between the measured local surface density near the Sun
with that predicted by maximum disk models, the detailed fit of the
terminal velocity curve by these models, and the need for substantial
baryonic mass in the inner Galaxy required by the bulge microlensing
observations (discussed below), all provide evidence that a maximum disk
solution in the Milky Way is approximately correct.

\section{Microlensing}

Microlensing observations provide important new constraints on the
Galactic mass distribution.  Several hundred microlensing events have
now been observed towards the Galactic bulge. These observations give
information about the integrated mass density towards the survey
fields as well as about the lens mass distribution. The most robust
observable is the total optical depth averaged over the observed
fields, $\tau$.  Early measurements gave surprisingly high values
$\tau_{-6}\simeq 2-4$ (Udalski \etal 1994, Alcock \etal 1997), where
$\tau_{-6}\equiv\tau/10^{-6}$.  Recently, Alcock et al.\ (2000a)
measured $\tau_{-6}=2.43^{+0.39}_{-0.38}$ for all sources from 99
events centered on $(l,b) = (2.68\deg, -3.35\deg)$, using a difference
image analysis (DIA) technique.  From this measurement they deduced
for the same direction $\tau=(3.23\pm0.5)\times10^{-6}$ for bulge
sources only. Finally, in a preliminary analysis of 52 clump giant
sources in 77 Macho fields, which do not suffer from blending
problems, Popowski \etal (2000) found a lower $\tau_{-6}=2.0\pm0.4$
centered on $(l,b)=(3.9\deg,-3.8\deg)$.

Axisymmetric Galactic models predict $\tau_{-6}\simeq 1-1.2$,
insufficient to explain the quoted optical depths (Kiraga \& Paczynski
1994, Evans 1994).  Models with a nearly end--on bar enhance $\tau$
because of the longer average line-of-sight from lens to source.  The
maximum effect occurs for $\phi\simeq \arctan(b/a)$ when $\tau_{\rm
  bar}/\tau_{\rm axi}\simeq (\sin2\phi)^{-1}\simeq 2$ for
$\phi=15\deg$ (Zhao \& Mao 1996).  In addition, $\tau$ increases with
the mass and the length of the bar/bulge.

\begin{figure}
\centerline{\epsfig{file=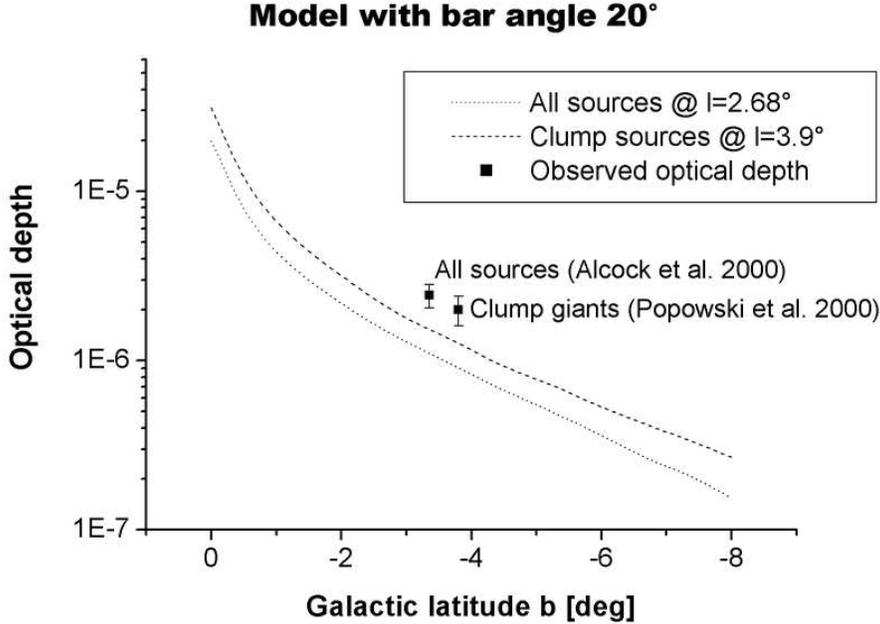,width=\hsize}}
\caption{Microlensing optical depths for the model of Fig.~1 at the
  longitudes of the newly published MACHO results, plotted as a
  function of galactic latitude. The points with error bars show the
  observed optical depths. The upper curve is for clump giant sources,
  the lower curve for all sources using a simple parametrization of
  the magnitude cut-off. From Bissantz \& Gerhard (2002). }
\end{figure}

However, models based on barred mass distributions derived from Milky
Way observations typically give $\tau_{-6}\simeq 1-2$ (e.g., Zhao,
Spergel \& Rich 1995, Stanek \etal 1997, Bissantz \etal 1997),
significantly less than most of the measured optical depths.  Figure 5
shows values of the optical depth for the new COBE bar model of
Bissantz \& Gerhard (2002) as function of latitude, at the central
longitude positions of the new microlensing measurements.  The mass
normalization of the disk and bulge in this model is calibrated by
assuming constant L-band mass-to-light ratio and by matching the
predicted gas flow velocities in a hydrodynamic simulation to the
Galactic terminal velocity curve; see \S\S 1,2. The numerical values
are $\tau_{-6}=1.1$ for all sources at the position of the DIA
measurement and $\tau_{-6}=1.27$ for clump giant sources at the
centroid position given by Popowski \etal (2000).  

Because the apparent magnitude distributions for clump giant stars
predicted by this model agree closely with those measured by Stanek
\etal (1997) -- see Fig.~2 -- this model gives a good approximation to
the distribution of microlensing {\sl sources}.  Changing the quoted
optical depths substantially is therefore hard unless the mass
distribution of the {\sl lenses} differs substantially from that of
the sources.

The model prediction for clump giant sources is within $1.8\sigma$ of
the Popowski \etal MACHO value. On the other hand, the recent DIA
value is still some $3.2\sigma$ away from the model prediction.  While
the NIR model prediction could be slightly increased if the
mass-to-light ratio were not spatially constant, this is only a $\sim
20\%$ effect since limited by the terminal velocity curve (Bissantz
\etal 1997).  Binney, Bissantz \& Gerhard (2000) used
general arguments to show that an optical depth for bulge sources as
large as that implied by the MACHO DIA measurement is very difficult
to reconcile with the Galactic rotation curve and local mass density,
even for a barred model and independent of whether mass follows light.
To illustrate this, the extra optical depth required would correspond
to an additional mass surface density towards the bulge of some $1500
\msun/\pc^2$ at the optimal location half-way to the bulge. This is
comparable to the luminous surface mass density in the NIR model (some
$3600 \msun/\pc^2$ but not optimally located, Bissantz \& Gerhard
2002). It will be important to check whether the DIA measurement could
still be significantly affected by blending.

Regardless of the resolution of this problem, these results have a
further important implication. If a model based on the maximal disk
assumption and calibrated with the terminal velocities still
underpredicts the observed bulge microlensing optical depths, we can
certainly not afford to lose a significant fraction of this mass to a
non-lensing CDM dark halo. Indeed, from the LMC microlensing
experiments (Alcock \etal 2000b) at most a fraction of the dark matter
halo contributes to microlensing. Thus the bulge microlensing results
independently argue strongly for a massive disk (see also Gerhard
2001 and Binney \& Evans 2001).

\section{The dark halo}

Determining the total size and mass of the Galactic halo is of
much interest, but has proved a difficult problem. The gas rotation
curve outside the solar radius is much less well known in the MW
than in other galaxies (see, e.g., Dehnen \& Binney 1998), and 
it does not extend beyond $\sim 20\kpc$. Estimates of the mass of
the outer halo have therefore principally relied on the measured
radial velocities and proper motions of the distant Galactic 
satellites and globular clusters (Little \& Tremaine 1987,
Kochanek 1996, Wilkinson \& Evans 1999). One problem has been
the sensitivity of the mass estimates to whether or not the
Leo I dwarf galaxy is assumed to be bound to the MW and included
in the analysis. This problem has only been overcome with the
availability of proper motion velocity measurements.

In the most recent analysis, Wilkinson \& Evans (1999) used radial
velocities of 27 objects at galactocentric distances greater than $20
\kpc$ and proper motions for six of these. They conducted a likelihood
analysis based on fitted parametric mass and number density models and
simple spherical anisotropic distribution functions for the
satellites. Within these model assumptions they determined
most probable values for
the total halo mass, size, and mass within $50\kpc$, as well as error
estimates based on the likelihood contours. The resulting total halo
mass within the cutoff radius (typically $\sim 200\kpc$) is $M_{\rm
  tot} \sim 1.9^{+3.6}_{-1.7} \times 10^{12}\msun$, and the mass
within $50\kpc$ is $\sim 5.4^{+0.2}_{-3.6} \times 10^{11}\msun$.
Values near the upper end of the range at $50\kpc$ are consistent with
a flat rotation curve with $v_c=220\kms$.  It is clear that the errors
on these estimates are still very considerable, mostly because of the
small sample size and the large proper motion errors. Wilkinson \&
Evans discuss how these estimates can be tightened in the future using
ground based radial velocities for a sample of $\sim 200$ BHB stars
and accurate satellite measurements of proper motions for the known
satellites.

The shape of the Galactic dark halo is constrained by measurements of
the distribution and kinematics of halo stars, data on the
distribution and kinematics of tidal streams, flaring of the gas
layer, and the comparison of microlensing optical depths along
different lines-of-sight. These constraints have recently been
reviewed by van der Marel (2001). While values of $q_h\gta 0.7$ for
the halo axial ratio appear most consistent with the data, it is hard
to rule out any axial ratio $0.4\lta q_h\lta 1.0$ from the present
observations.  Perhaps the most promising method is that based on
tidal streams. At present Ibata \etal (2001) base their analysis of
the Sagittarius stream on only 38 Carbon stars, inferring from their
apparent distribution on a great circle on the sky and a comparison
with numerical simulations that the dark halo must be nearly
spherical, with $q_h\gta 0.7$. Similar, larger datasets would clearly
be very valuable.

%\begin{acknowledgements}
%\end{acknowledgements}

\end{article}

\end{document}